\documentclass[10pt,conference]{IEEEtran} 
\IEEEoverridecommandlockouts
\usepackage{cite}
\usepackage{amsmath,amssymb,amsfonts}
\usepackage{algorithmic}
\usepackage{graphicx}
\usepackage{textcomp}
\usepackage{subcaption}
\usepackage{xcolor}
\usepackage{tikz}
\usetikzlibrary{shapes.geometric, arrows, positioning, decorations.pathreplacing}

\usetikzlibrary{shapes.geometric, arrows}
\usepackage{algorithm}
\usepackage{algorithmic}

\definecolor{amber}{rgb}{1.0, 0.75, 0.0}
\definecolor{almond}{rgb}{0.94, 0.87, 0.8}
\definecolor{blond}{rgb}{0.98, 0.94, 0.75}
\definecolor{cornflowerblue}{rgb}{0.39, 0.58, 0.93}
\definecolor{lavenderblue}{rgb}{0.8, 0.8, 1.0}
\definecolor{lightskyblue}{rgb}{0.53, 0.81, 0.98}
\definecolor{lime(web)(x11green)}{rgb}{0.0, 1.0, 0.0}
\definecolor{lime(colorwheel)}{rgb}{0.75, 1.0, 0.0}
\definecolor{persianpink}{rgb}{0.97, 0.5, 0.75}
\definecolor{mistyrose}{rgb}{1.0, 0.89, 0.88}
\definecolor{ticklemepink}{rgb}{0.99, 0.54, 0.67}
\definecolor{salmonpink}{rgb}{1.0, 0.57, 0.64}
\definecolor{richbrilliantlavender}{rgb}{0.95, 0.65, 1.0}
\definecolor{pink}{rgb}{1.0, 0.75, 0.8}
\definecolor{cadetgrey}{rgb}{0.57, 0.64, 0.69}
\definecolor{darkpastelblue}{rgb}{0.47, 0.62, 0.8}

\def\BibTeX{{\rm B\kern-.05em{\sc i\kern-.025em b}\kern-.08em
    T\kern-.1667em\lower.7ex\hbox{E}\kern-.125emX}}
\begin{document}

\title{Adaptive Communication Through Exploiting RIS, SSK, and CIM for Improved Reliability and Efficiency}
\author{Ferhat Bayar \IEEEmembership{Student Member, IEEE}, Onur Salan, Erdogan Aydin and Haci Ilhan, \IEEEmembership{Senior Member, IEEE}
 \thanks{F. Bayar is with the Scientific and Technological Research Council of Türkiye (TUBITAK), B{I}LGEM, Kocaeli, Turkey (e-mail: ferhat.bayar@tubitak.gov.tr) (Corresponding author: Ferhat Bayar).}
 \thanks{O. Salan is with Communications and Signal Processing Research (HISAR) Laboratory, T{U}B{I}TAK B{I}LGEM, Kocaeli, Turkey (e-mail: onur.salan@tubitak.gov.tr).}
  \thanks{E. Aydin is with the Department of Electrical and Electronics Engineering, Istanbul Medeniyet University, Istanbul 34857, Turkey (e-mail: erdogan.aydin@medeniyet.edu.tr).}
 \thanks{H. Ilhan is with the Department of Electronics and Communications Engineering, Yildiz Technical University, Istanbul, 34220, Turkey, (e-mail: ilhanh@yildiz.edu.tr).}

}

\maketitle

\begin{abstract}
In this paper, we present a novel communication system model that integrates reconfigurable intelligent surfaces (RIS), spatial shift keying (SSK), and code index modulation (CIM) based on Hadamard coding called RIS based transmit SSK-CIM (RIS-CIM-TSSK). By leveraging RIS, the system adapts rapidly to dynamic environments, enhancing error rates and overall reliability. SSK facilitates the transmission of additional passive information while eliminating the need for multiple radio frequency (RF) chains, thereby reducing complexity. CIM enhances passive information transmission through frequency domain spreading, which may increase signal obfuscation. This proposed scheme not only improves energy efficiency but also offers a robust solution for reliable communication in modern wireless networks, paving the way for smarter and more adaptable implementations. We consider a suboptimal, low-complexity detector for the proposed scheme and also address the blind case for phase adjustment of the RIS. Finally, we present the simulation results for the proposed system model across various configurations, including different numbers of receive and transmit antennas, varying reflecting elements of the RIS, and different code lengths.
\end{abstract}

\section{Introduction}
The rapid advancement of wireless communication technologies has significantly transformed the landscape of connectivity, paving the way for next-generation systems such as five generation (5G) and the upcoming sixth generation (6G). These advancements aim to support an ever-increasing demand for high data rates, low latency, and reliable connectivity across various applications, from smart cities to the Internet of Things (IoT). As we look forward to 5G and beyond, expectations in wireless communication continue to evolve. 5G is anticipated to deliver unprecedented speeds, ultra-reliable low-latency communication, and massive connectivity for IoT devices. In contrast, 6G is expected to push these boundaries even further, targeting data rates exceeding 100 gigabit per seconds (Gbps), sub-millisecond latencies, and advanced applications such as holographic communication and intelligent networks that can autonomously adapt to user needs. The integration of reconfigurable intelligent surfaces (RIS), code index modulation (CIM), and space shift keying (SSK) will play a crucial role in realizing these expectations, enabling efficient spectrum utilization and enhancing overall system performance.

RIS represents a transformative technology in wireless communications, designed to enhance signal quality and network performance by dynamically adjusting the phase shifts of passive elements to optimize channel conditions \cite{10380596}. When combined with advanced modulation techniques, such as CIM and spatial modulation (SM), RIS and media based modulation (MBM) can further improve system performance \cite{ozden2023reconfigurable}. The CIM particularly within the framework of code division multiple access (CDMA), introduces a new dimension to data transmission. CIM, which transmits information by selecting from a predefined codebook of codewords, benefits from RIS by optimizing channel conditions for more effective codeword selection. By encoding information into the indices of code-words rather than solely using amplitude or phase, CIM provides a novel means to achieve higher spectral efficiency and greater resilience against interference \cite{7317808}. This technique is particularly well-suited for the high user densities anticipated in future networks, allowing for simultaneous transmissions with minimal degradation in performance. In study \cite{8792959} authors introduce joint code-frequency-index modulation (CFIM) by considering code and frequency domains for index-modulation (IM). Similarly, SSK, which encodes information through the selection of specific antennas, can leverage RIS to improve energy efficiency and reduce system complexity. In \cite{canbilen2020reconfigurable} the authors explore the potential of integrating RIS with index modulation techniques as a novel approach for future 6G wireless systems. This work highlights how this paradigm can enhance system capacity and spectral efficiency beyond traditional multiple-input multiple-output (MIMO) technologies, paving the way for advanced communication solutions. 

\begin{figure*}
\centering
\includegraphics[width=\textwidth]{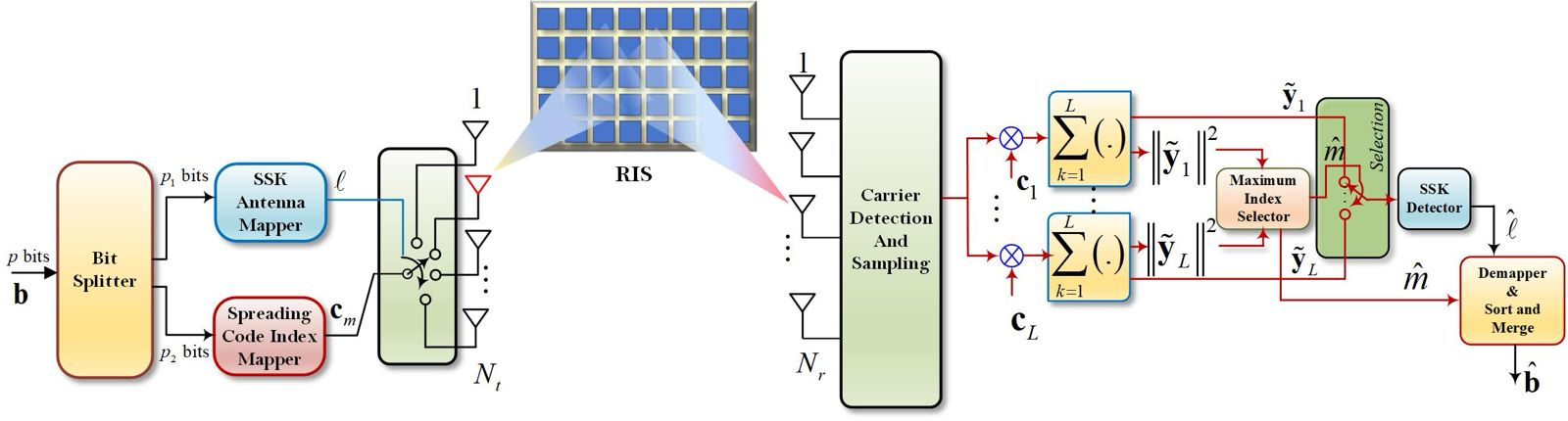}
\caption{System model of RIS-CIM-TSSK }
\label{fig:system_model}
\end{figure*}

SSK technique complements the benefits offered by RIS and CIM, forming a synergistic relationship that can address the demands of future wireless communication systems. In study \cite{cogen2024reconfigurable} the authors investigate the integration of RIS with CIM to enhance the performance of single-input single-output (SISO) communication systems. This work demonstrates how RIS can significantly improve data transmission rates and reliability in high-rate scenarios, providing a novel approach to optimizing communication performance. In \cite{ozden2024novel}, the authors present an innovative approach that combines RIS with code index modulation and receive spatial modulation (RSM). This work examines how the integration of these technologies can enhance communication system performance, improving data rates and transmission reliability in various scenarios. The study \cite{singh2023indexed} explores the innovative application of RIS to enhance indexed multiple access (IMA) systems. The authors focus on the concept of reflection tuning, demonstrating how RIS can dynamically adjust reflection coefficients to optimize signal quality and user connectivity. In study \cite{bayar2024space}, the integration of space-time block coding (STBC) with RIS in RSM systems. The authors propose a novel framework that leverages STBC to enhance the performance of RIS-aided communication by improving signal reliability and diversity. The study \cite{li2024exploiting} investigates the integration of RIS in SM systems by introducing an activation mode index. The authors explore how this index can enhance the SM technique by enabling more efficient utilization of the RIS, thereby improving signal quality and system performance. 

In this study, we will explore the potential of RIS, CIM, and SSK in shaping the future of wireless communication. We will propose RIS-CIM-TSSK system and examine their individual contributions and collaborative capabilities in meeting the demands of 5G and 6G networks, ultimately laying the groundwork for a more intelligent , robust and efficient communication systems. Also we will present correlation based low complexity detector for proposed system in order to provide flexible approach. 

\subsection{Notations}
In this paper , matrices and vectors are shown in boldface uppercase and boldface lowercase letter, respectively. $\left(\cdot\right)^*$, $\left(\cdot\right)^T$, and $\left(\cdot\right)^H$ define complex conjugation, transpose, and Hermitian transpose, respectively.  

\subsection{Paper Organization}
The remainder of this paper is organized as follows. Section II outlines the RIS-CIM-TSSK schemes with a low complexity (LC) receiver and defines the receiver structure. The Monte Carlo simulation results are presented in Section III. Finally, Section IV concludes the paper.

\section{System Model}
Fig. \ref{fig:system_model} depicts a system model of RIS-TSSK-CIM in MIMO system, where transmitter is equipped with $N_t$ antennas, the receiver with $N_r$ antennas and the RIS with $N$ passive reflecting elements. The RIS is used as a relay-mode. Note that there is no line of sight (LOS) between the receiver and the transmitter. Transmission from the receiver to the sender only occurs through RIS.

For the RIS-CIM-TSSK scheme, the incoming $p = \log_2({N_t})+\log_2({L})$ information bits are portioned and processed into two sub groups. The first part, which contains $p_1 = \log_2({N_t})$ bits, selects the transmitter antenna index $l$. The other $p_2 = \log_2({L})$ bits are assigned to transmitted code-word index $m$. By considering both the two passive information, the data rate of RIS-TSSK-CIM system in bits per channel use (bpcu) can be expressed as :
\begin{equation}
p = p_1 + p_2 = \log_2 N_t + \log_2 L. 
\end{equation}

Particularly, we constructed a spreading codebook using Hadamard codes, denoted as $\mathbf{C} = \{ \mathbf{c}_1, \ldots, \mathbf{c}_{L} \} $ where $\mathbf{c}_m \in \mathbb{R}^{F \times 1}$ for a given code length $F$ and a number of codes $L$. Specifically, we set the code length and the number of codes equal, such that $F=L$.  To ensure that the codewords have a unit norm, we normalize each codeword by multiplying by $\frac{1}{\sqrt{L}}$ . This normalization step is crucial for maintaining consistent power levels across the codewords. Based on the passive information of $\log_2 L$ bits, the selection of the code $\mathbf{c}_m$ is utilized to spread the base-band signal. Note that we assume that the channel remains stationary during each code chip, implying that it does not change throughout the duration of the code length.

The wireless channel $\mathbf{G}$ between transmitter and RIS for the $u$th chip can be represented as follows:
\begin{equation}
    \mathbf{G}^u = \begin{bmatrix}
       g_{1,1}^u & g_{1,2}^u & \cdots & g_{1,N}^u 
       \\
       g_{2,1}^u & g_{2,2}^u & \cdots & g_{2,N}^u 
       \\
       \vdots & \vdots & \vdots & \vdots \\ 
       g_{N_t,1}^u & g_{N_t,2}^u & \cdots & g_{N_t,N}^u
    \end{bmatrix}_{N_t\times N}
    \label{channel_matrix_bs_to_ris}
\end{equation}
where $g_{l,i}^u$ is channel entity between $k$th transmit antenna and $i$th reflector element for the $u$th chip, it is characterized as following: 
\begin{equation}
g_{k,i}^u=\beta_{k,i}^ue^{-j\theta_{k,i}^u} \quad k= 1, 2, \cdots, \textit N_t, \quad i= 1, 2, \cdots, \textit N.
\label{bs_to_ris_channel}
\end{equation}
Here, $\theta_{k,i}^u$ is the channel phase induced by the $i$th reflector at the $k$th transmit antenna index and $\beta^u_{k,i}$ is channel fading coefficients between the $i$th reflector and $k$th transmit antenna which follows a Rayleigh distribution. Since the selection of the $l$th transmitting antenna is based on the incoming bit data sequence, the channel between the transmitter and the RIS will be represented in the following form:

\begin{equation}
    \mathbf{g}_{l}^u = \begin{bmatrix}
       g_{1,l}^u & g_{2,l}^u & \cdots & g_{N,l}^u 
    \end{bmatrix}
    \label{vector_bs_to_ris}_{1\times N}
\end{equation}

Similarly, the wireless channel matrix $\mathbf{H}$ with $N\times N_r$ dimension between RIS and receiver for the $u$th chip can be shown as follows:
\begin{equation}
    \mathbf{H}^u = \begin{bmatrix}
       \kappa_{1,1}^u & \kappa_{1,2}^u & \cdots & \kappa_{1,N_r}^u 
       \\
       \kappa_{2,1}^u & \kappa_{2,2}^u & \cdots & \kappa_{2,N_r}^u 
       \\
       \vdots & \vdots & \vdots & \vdots \\ 
       \kappa_{N,1}^u & \kappa_{N,2}^u & \cdots & \kappa_{N,N_r}^u
    \end{bmatrix}_{N\times N_r}
    \label{Matrix for RIStoReceiver}
\end{equation}
where $\kappa_{n,i}^u$ the wireless fading channel entity between $n$th receive antenna and $i$th reflector element is characterized by
\begin{equation}
\kappa_{i,n}^u=\alpha^u_{i,n}e^{-j\Psi^u{i,n}} \quad n= 1, 2, \cdots, \textit N_r,
\label{ris_to_destination_channel}
\end{equation}
where $\Psi^u_{n,i}$ is the channel phase induced by the $i$th reflector at the $n$th receive antenna and $\alpha_{n,i}^u$ is channel fading coefficients between the $i$th reflector and $n$th receive antenna follows Rayleigh distribution. The reflection coefficient of the \(i\)-th RIS element is expressed as \(\Phi^u_i = e^{j\phi^u_i}\), where \(\Phi^u_i \in [0, 2\pi)\) denotes the phase shift for \(i = 1, \ldots, N\). Most studies assume that the channel phase \(\Phi^u_i\) is perfectly known at the RIS, allowing the reflector to adjust to the RIS phase \(\Phi^u_i\) without error, represented as \(\Phi^u_i = \theta^u_{l,i} + \Psi^u_{n,i}\). In this work, we also consider a blind RIS control scenario, in which the RIS operates without access to instantaneous or statistical CSI. Specifically, we set all RIS phase shifts to zero, i.e., $\phi^u_i = 0$, resulting in uniform reflection coefficients across all elements, $\Phi^u_i = 1$. This naive approach assumes no optimization at the RIS side and reflects a low-complexity baseline, suitable for evaluating the performance gain achievable through RIS phase adaptation.

Additionally, we can represent the RIS phases with the dimention $N \times N$ in matrix form as follows:
\begin{equation}
   \mathbf{\Phi} = \text{diag}(e^{j\phi^u_1}, e^{j\phi^u_2}, \ldots, e^{j\phi^u_N})
   \label{Phase matrix on RIS}
\end{equation}

The received base-band noisy signal $\mathbf{Y}$ with the dimension $N_r \times L$ is represented as:
\begin{equation}
\textbf{Y}= \mathbf{H}^{T}\mathbf{\Phi}\mathbf{g}_{l}^T \textbf{c}_{m} + \textbf{W},
\label{received_sginal}
\end{equation}
where $\mathbf{W}$ stands for additive white Gaussian noise (AWGN) with $N_r\times L$. 


\subsection{Suboptimal Low Complexity (LC) Detector}
This section presents a low-complexity sub-optimal receiver for the RIS-CIM-TSSK system. This receive design follows a greedy approach, iterating through the codebook to identify the vector that yields the highest correlation with the received signal. The selected code index can be recovered with the following maximization method \cite{li2023ris}:
\begin{equation}
\hat{m} = \arg \max_{\mathbf{c} \in \mathbf{C}} \biggl\{\Big\vert\Big\vert  \mathbf{Y}\mathbf{c}^{T} \Big\vert\Big\vert^2 \biggl\}
\label{code_detection}
\end{equation}
where $\hat{m}$  represents the estimated code-book index, corresponding to the code vector selected for optimal performance. Once the code index is determined, the next step is to recover the active transmitter index using the following criterion:
\begin{equation}
\hat{l} = \arg \min_{l = 1, 2, \ldots, N_t} \Big\vert\Big\vert \textbf{Y}\textbf{c}_{\hat{m}}^T  - \mathbf{H}^T \mathbf{\Phi}\mathbf{g}_{l}^T \Big\vert\Big\vert^2.
\label{antenna_detection}
\end{equation}
where, $\hat{l}$ denotes the estimated index of the active transmitter. The recovery process begins with the estimation of the code index, as detailed in (\ref{code_detection}), followed by the determination of the selected antenna index in (\ref{antenna_detection}). This two-step approach effectively identifies both the optimal code and the corresponding transmitter, ensuring efficient signal detection in the RIS-CIM-TSSK framework.

\vspace{-1mm}  
\section{Simulation Results}
\begin{figure}
   \begin{subfigure}[a]{0.49\textwidth}
     \centering
     \includegraphics[width=\textwidth]{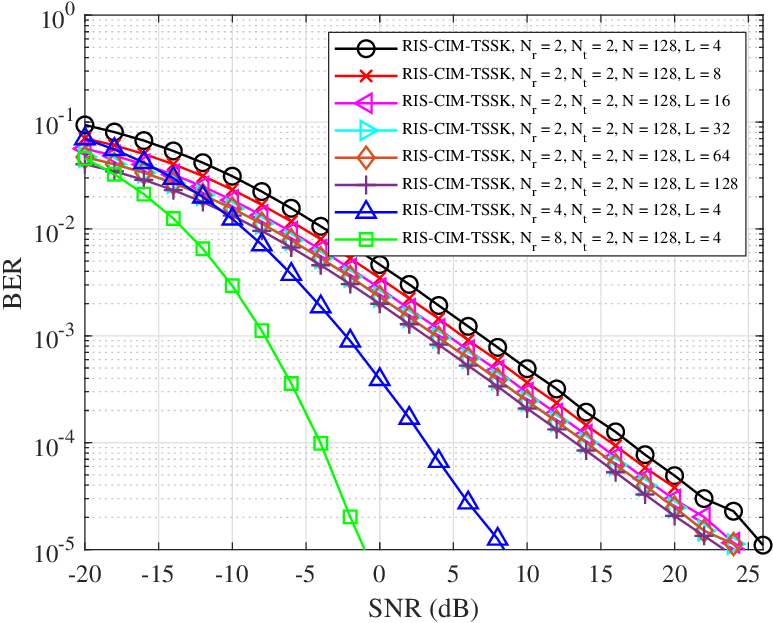}
     \caption{BER performance of the RIS-based CIM-TSSK schemes with different $L$ and $N_r$ ($N_t=2$, $N= 128$).}\label{Fig:Varying L and Nr for RIS-CIM-TSSK}
   \end{subfigure}
   \hfill
   \begin{subfigure}[a]{0.49\textwidth}
     \centering
     \includegraphics[width=\textwidth]{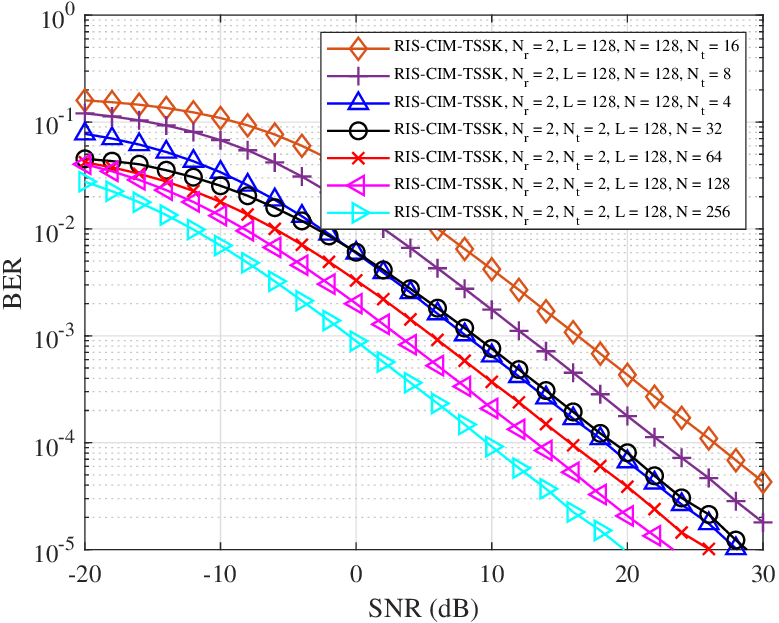}
     \caption{BER performance of the RIS-based SM schemes with different $N$ and $N_t$ ($L=128$, $N_r=2$).}\label{Fig:Varying N and Nt for RIS-CIM-TSSK}
   \end{subfigure}
   \caption{BER performance of RIS-CIM-TSSK with varying numbers of reflectors, as well as transmit and receive antennas.}
\end{figure}

Fig. \ref{Fig:Varying L and Nr for RIS-CIM-TSSK} represents as the code length $L$ increases (from $4$ to $128$), the bit error rate (BER) generally slightly improves across all signal-to-noise ratio (SNR) values. This situation suggests that the elimination of channel phases by the RIS adversely affects its performance in relay communication. However, it can also be interpreted that increasing the length of the code enhances the overall system diversity. For instance, the BER for $L = 4$ is higher compared to $L = 128$ at the same SNR levels. Fig. \ref{Fig:Varying L and Nr for RIS-CIM-TSSK} also shows the effect of number of receive antennas on BER performance of proposed scheme. It is clearly seen that the system's performance can also be positively affected by the number of receive antennas. For a fixed code length $L$ and transmit antenna $N_t$, increasing the number of receive antennas can enhance system performance. For example, results with $N_r = 8$ and $N_t = 2$ tend to show better BER performance than configurations with fewer receive antennas. This is evident when comparing BERs for configurations with different numbers of receive antennas (e.g., $N_r = 2$ vs. $N_r = 4$ and $N_r = 8$). The reason for this situation can be attributed to the increase in the number of antennas, which enhances diversity at the receiver and positively impacts the BER.

Fig. \ref{Fig:Varying N and Nt for RIS-CIM-TSSK} demonstrates that increasing the number of RIS elements consistently improves the BER performance across the SNR range. Higher passive RIS elements lead to better error mitigation and lower BER. On the other hand, increasing the number of transmitters does not always guarantee improved performance; it can sometimes lead to higher BER due to the challenges associated with distinguishing between transmitter indices at the receiver.

\begin{figure}
   \begin{subfigure}[a]{0.49\textwidth}
     \centering
     \includegraphics[width=\textwidth]{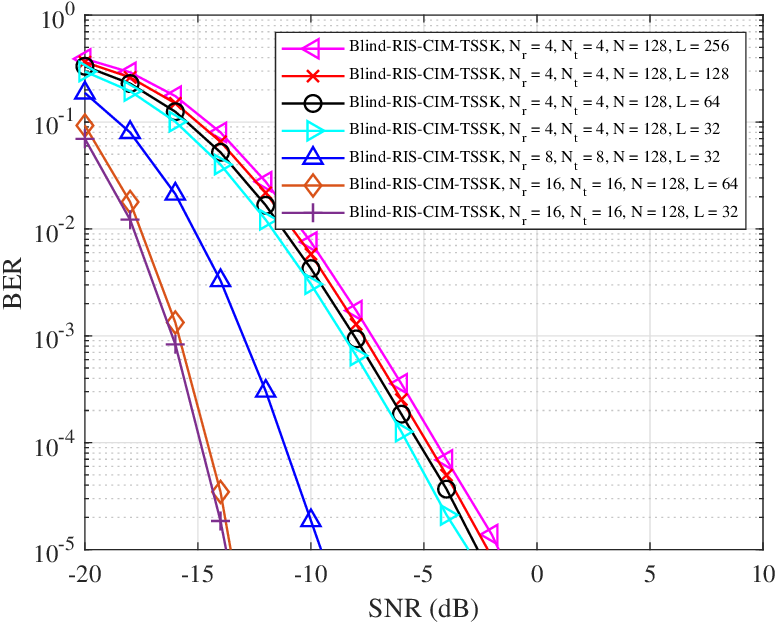}
     \caption{BER performance of Blind RIS-CIM-SKK for varying number of pbcu.}\label{Fig:Varying bpcu for Blind-RIS-CIM-SSK}
   \end{subfigure}
   \hfill
   \begin{subfigure}[a]{0.49\textwidth}
     \centering
     \includegraphics[width=\textwidth]{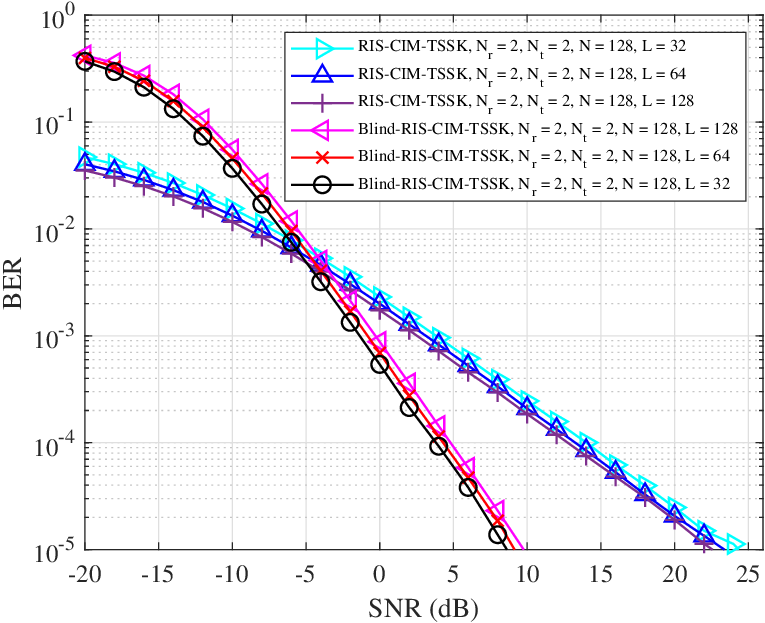}
     \caption{BER performance comparison of the RIS-CIM-SSK and Blind-RIS-CIM-SSK with different $L$ ($N_t=2$, $N_{r}=2$ and $N=128$).}\label{Fig:Varying L for RIS-based CIM-SSK and Blind RIS-based CIM-SSK}
   \end{subfigure}
   \caption{BER performance of RIS-CIM-TSSK and Blind-RIS-CIM-TSSK for varying system settings.}
\end{figure}

Fig. \ref{Fig:Varying bpcu for Blind-RIS-CIM-SSK} demonstrates the performance of proposed scheme across various bpcu. In first scenario each system configuration utilized $4$ transmit antennas and $4$ receive antennas, with the number of passive elements in the RIS set to $N=128$. The blind case indicates that the RIS phases were adjusted to $1$. The configurations explored include different code lengths: $32$, $64$, $128$, and $256$, affecting the overall bit transmission capabilities. For example, with a code length of $64$ and $8$ bpcu, the system exhibited BER that improved significantly as the SNR increased, transitioning from 0.3327 at -$20$ dB to nearly zero at $10$ dB. Conversely, when the code length was increased to $128$ with $9$ bpcu, the initial BER was higher at $0.3640$ at $-20$ dB, but it also showed improvement with increasing SNR, stabilizing at zero for higher SNR values. It is expected the seen that increasing the bpcu generally led to lower BER values.  In second scenario also effect of number of receive antenna is presented as well as code length. Despite the fact that increasing the bpcu negatively impacts system performance, it is evident that the rise in the number of receive antennas significantly enhances the BER performance in relation to this parameter.

Fig. \ref{Fig:Varying L for RIS-based CIM-SSK and Blind RIS-based CIM-SSK} presents comparison of Blind RIS-CIM-TSSK and non-blind (optimized phase) RIS-CIM-TSSK. Obtanied results show that particularly at low SNR levels, the blind RIS results in lower BER compared to optimized RIS settings. The blind RIS configurations demonstrate a degree of robustness, as they manage to reduce BER significantly with increasing SNR, albeit from a higher starting point. This suggests that while blind systems may be less efficient initially, they can still provide effective error performance through dynamic adjustments. The need for blind RIS configuration presents a trade-off between flexibility and performance. Blind RIS systems may require higher SNR thresholds to achieve comparable BER performance with non-blind configurations. Both blind and non-blind systems show a clear trend where BER decreases as SNR increases. However, the performance gap at lower SNR levels underscores the advantages of utilizing channel knowledge in non-blind configurations. For practical applications, incorporating elements of both approaches may be beneficial. Strategies that allow for some degree of channel knowledge in blind configurations could further enhance performance, especially in challenging environments with variable channel conditions.

\begin{figure}[!t]
    \centering
    \includegraphics[width=1\linewidth]{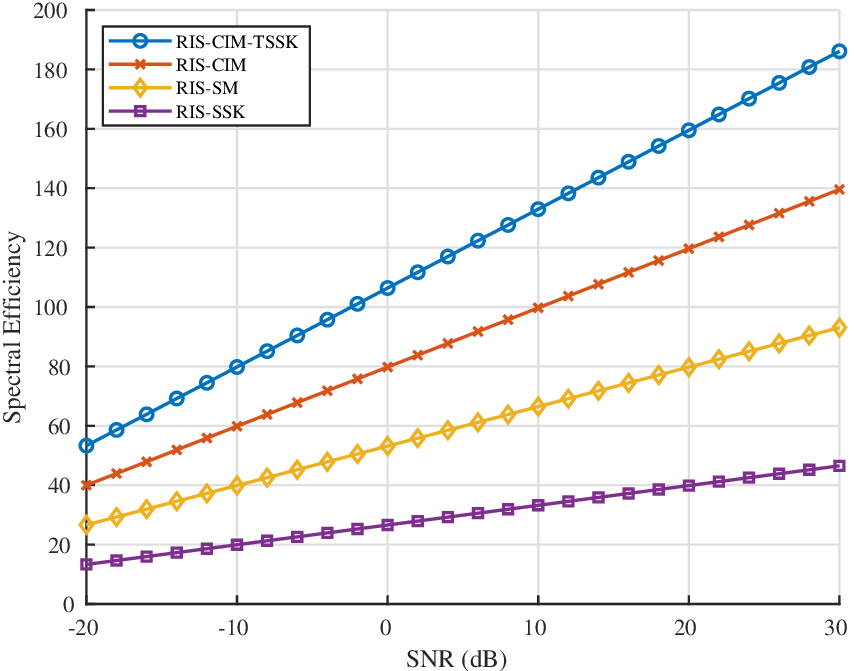}
    \caption{Spectral efficiency versus SNR for RIS-CIM-TSSK, RIS-CIM, RIS-SM, and RIS-SSK schemes.}
    \label{Fig:spectral_efficiency}
\end{figure}

\noindent
Fig. \ref{Fig:spectral_efficiency} illustrates the spectral efficiency performance of four different RIS-assisted modulation schemes across a range of SNR values. The compared schemes include the proposed RIS-CIM-TSSK, conventional RIS-CIM, RIS-SM, and RIS-SSK. The system under consideration includes $N_t=4$ transmit antennas, $N=128$ passive reflecting elements, and number of codes $L=64$. The RIS-SM scheme uses QPSK ($M= 4$) modulation. As observed, the RIS-CIM-TSSK scheme consistently achieves higher spectral efficiency compared to the other schemes. This improvement is primarily due to its ability to jointly exploit via transmit antenna selection, codeword indexing. In contrast, the RIS-CIM scheme utilizes only codeword indexing without spatial modulation, resulting in lower performance. The RIS-SM scheme combines antenna indexing with conventional modulation, but lacks codebook-based indexing, which limits its throughput. Lastly, RIS-SSK shows the lowest spectral efficiency, as it relies solely on antenna indexing with no symbol or codeword modulation.

\section{Conclusion}
In conclusion, the integration of RIS, CIM, and SSK represents a significant advancement in the evolution of wireless communication technologies. The proposed system model leverages RIS to facilitate smart implementation, enabling rapid adaptation to changing environments and improving error rates significantly. By integrating SSK, the system efficiently transmits additional passive information without the need for multiple RF chains, thus reducing complexity and resource requirements. Furthermore, CIM improves passive information transmission by spreading the signal across the frequency domain, which may inherently increase signal obfuscation. While this characteristic could potentially make signal detection more difficult for unintended receivers, it is important to note that a formal security analysis such as the definition of a threat model or evaluation of secrecy metrics is not within the scope of this work. Future research could explore these aspects in more detail. Together, these elements create a robust communication framework that excels in adaptability, efficiency, and system-level performance.
\vspace{-1mm}  

 \section*{Acknowledgement}
This work was supported by the Scientific and Technological Research Council of Türkiye (TUBITAK) under the Project-$123$E$513$.
\vspace{-6mm}  

\vspace{12pt}

\bibliographystyle{IEEEtran}
\bibliography{IEEEabrv,refernces}

\end{document}